\documentclass[reprint,superscriptaddress,prb,showpacs]{revtex4-1}

\usepackage{graphicx}
\usepackage{epstopdf}
\usepackage[ansinew]{inputenc}
\usepackage{array}
\usepackage{color}
\usepackage{amsmath}
\usepackage{amsxtra}
\usepackage{amstext}
\usepackage{amssymb}
\usepackage{latexsym}
\usepackage{dsfont}

\begin{document}
\title{Local Spectroscopy of the Electrically Tunable Band Gap in Trilayer Graphene}
\author{Matthew Yankowitz}
\affiliation{Physics Department, University of Arizona, 1118 E 4th Street, Tucson, Arizona 85721, USA}
\author{Fenglin Wang}
\affiliation{Department of Physics and Astronomy, University of California, Riverside, California 92521, USA}
\author{Chun Ning Lau}
\affiliation{Department of Physics and Astronomy, University of California, Riverside, California 92521, USA}
\author{Brian J. LeRoy}
\email{leroy@physics.arizona.edu}
\affiliation{Physics Department, University of Arizona, 1118 E 4th Street, Tucson, Arizona 85721, USA}
\date{\today}


\begin{abstract}
The stacking order in trilayer graphene plays a critical role in determining the existence of an electric field tunable band gap.  We present spatially-resolved tunneling spectroscopy measurements of dual gated Bernal (ABA) and rhombohedral (ABC) stacked trilayer graphene devices.  We demonstrate that while ABA trilayer graphene remains metallic, ABC trilayer graphene exhibits a widely tunable band gap as a function of electric field.  However, we find that charged impurities in the underlying substrate cause substantial spatial fluctuations of the gap size.  Our work elucidates the microscopic behavior of trilayer graphene and its consequences for macroscopic devices.
\end{abstract}

\maketitle

Graphene has great potential to be used in novel electronics applications ~\cite{Kim2012} due to its extraordinarily rich physical properties ~\cite{Zhang2005,Novoselov2005,CastroNeto2009,DasSarma2011}.  Many of these applications require inducing a sizable band gap without sacrificing its high intrinsic carrier mobility ~\cite{Schweirz2012}. While this has thus far not been achieved in single layer graphene, multilayer graphene has the possibility of an electric field tunable band gap.  Trilayer graphene exhibits two natural stacking orders, Bernal and rhombohedral. The more commonly found Bernal-stacked trilayer graphene is not expected to exhibit a significant field tunable band gap due to its mirror symmetry ~\cite{Guinea2006,Aoki2007,Koshino2009,Peeters2009a,Peeters2009b,Koshino2010,Kumar2011,Wu2011,Tang2011}, and experimental evidence thus far has supported this ~\cite{Craciun2009,JH2011,Henriksen2012}.  Its zero-field low energy band structure behaves roughly like a decoupled stack of Bernal-stacked bilayer graphene and single layer graphene ~\cite{Guinea2006,Aoki2007,Koshino2009,Peeters2009a,Peeters2009b,Koshino2010,Kumar2011,Wu2011,Tang2011}.  However, rhombohedrally-stacked trilayer graphene behaves differently in an electric field owing to its lack of mirror symmetry.  In this case, the low energy bands mimic a single sheet of Bernal-stacked bilayer graphene, and as such, a large field tunable band gap is expected in ABC trilayer graphene ~\cite{Aoki2007,Koshino2010,Kumar2011,Wu2011,Tang2011,Peeters2010,MacDonald2010}.  Prior work has explored the zero-field band structure of this material via electrical transport measurements and found a small band gap attributed to many-body effects ~\cite{Lau2011}.  The magnitude of the band gap in a tunable electric field has also been studied optically ~\cite{Heinz2011}.  Recent transport measurements have explored the global behavior of both stacking orders in an electric field ~\cite{Zhu2013}.  However, direct spectroscopic confirmation of the electronic properties of ABC trilayer with independent control of the Fermi energy and electric field has thus far been lacking. 

In this paper, we present scanning tunneling microscopy (STM) and scanning tunneling spectroscopy (STS) measurements of dual gated exfoliated samples of both ABA- and ABC-stacked trilayer graphene on Si/SiO$_2$ substrates.  Our results provide a direct understanding of the behavior of the band gap of ABC trilayer graphene in an electric field, and illuminate the effects of local disorder on its electronic properties.

\begin{figure}[h]
\includegraphics[width=8.6cm]{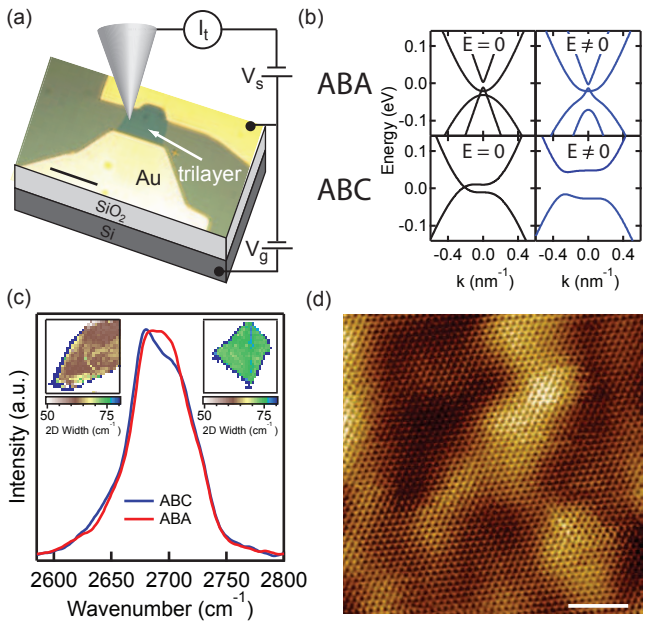} 
\caption{(Color online) Graphene device schematic, Raman spectroscopy and sample togography. (a) Schematic of the measurement setup showing the STM tip and an optical microscope image of one of the measured samples.  Scale bar is 20 $\mu$m.  (b) Band structure for ABA and ABC trilayer graphene with no electric field and a moderately sized electric field.  To first order, the low energy bands of ABA trilayer graphene are the superposition of the single-layer (roughly linear) and Bernal-stacked bilayer (roughly quadratic) graphene band structures, while the low energy bands of ABC trilayer graphene are roughly cubic.  (c) Raman spectroscopy of both stacking orders of trilayer graphene. Left upper inset: map of the FWHM of the 2D peak for an ABA-stacked flake.  Right upper inset: map of the FWHM of the 2D peak for an ABC-stacked flake. (d) STM topography image of ABC trilayer graphene showing the triangular lattice.  Scale bar is 2 nm.  Imaging parameters were sample voltage -200 mV and tunneling current 100 pA.}
\label{fig:schematic}
\end{figure}

Trilayer graphene flakes were mechanically exfoliated on 300 nm thick SiO$_2$ thermally grown on heavily doped Si substrates.  Cr/Au electrodes were written using electron beam lithography.  The devices were annealed at 350 $^{\circ}$C for 2 h in a mixture of argon and hydrogen and then at 300 $^{\circ}$C for 1 h in air before being transferred to the ultrahigh vacuum low-temperature STM for topographic and spectroscopic measurements.  Figure 1(a) shows a schematic diagram of the measurement set-up used for imaging and spectroscopy of the trilayer graphene flakes.  The layer number and stacking order of each flake was characterized using confocal Raman spectroscopy mapping prior to electrode deposition.  Representative Raman signals, taken with a 532 nm wavelength laser, are shown in  Fig.~\ref{fig:schematic}(c) for both stacking orders.  ABA stacked trilayers show a nearly symmetric 2D peak while ABC stacked trilayers show a wider and asymmetric peak ~\cite{Lui2010,Lau2011,Heinz2011,Zaliznyak2011,Cong2011}.  Only flakes exhibiting homogenous stacking order were used to ensure that local STM measurements were made on a known stacking order.  The insets to Fig.~\ref{fig:schematic}(c) show spatial maps of the FWHM of the 2D peaks for an ABA- and ABC-stacked sample, indicating that the flakes have a uniform stacking order.  

All the STM measurements were performed in ultrahigh vacuum at a temperature of 4.5 K.  dI/dV measurements were acquired by turning off the feedback circuit and adding a small (5 mV) a.c. voltage at 563 Hz to the sample voltage.  The current was measured by lock-in detection.  Topographically, we measure a triangular lattice in both types of trilayer graphene samples, as illustrated for ABC trilayer graphene in Fig.~\ref{fig:schematic}(d), which is consistent with the expected topography for a multilayer graphene sample.

\begin{figure}[h]
\newpage
\includegraphics[width=8.6cm]{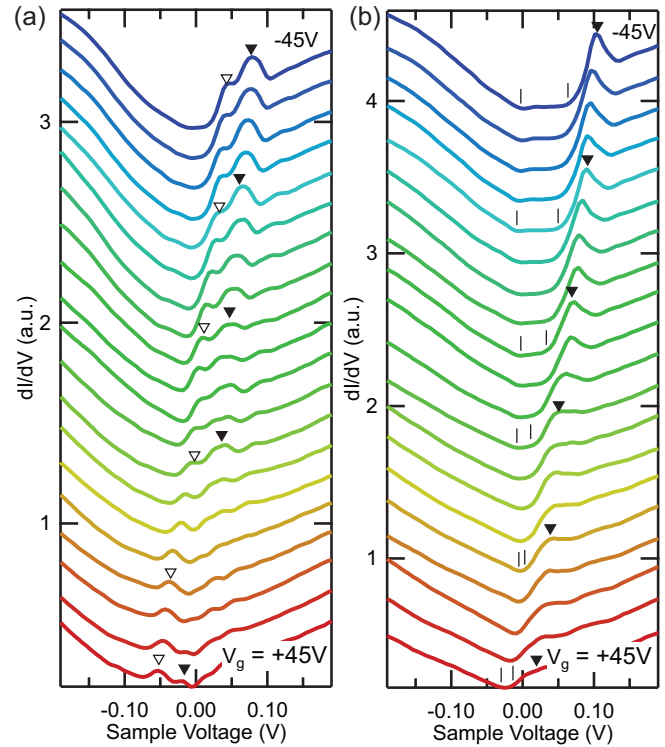} 
\caption{(Color online) Density of states of ABA and ABC trilayer graphene at varying gate voltage. (a)  Experimental dI/dV curves for ABA trilayer graphene taken in regular intervals of back gate voltage between +45 V and -45 V in steps of -4.74 V.  Black arrows indicate the location of the bilayer-like conduction band edge.  White arrows indicate the location of the bilayer-like valence band edge.   (b) Experimental dI/dV curves for ABC trilayer graphene for back gate voltages ranging from +45 V and -45 V in steps of -4.74 V.  Black arrows indicate the location of the conduction band van Hove singularity.  Black tick marks represent the band gap edges.  All curves are the average of 64 measurements taken over a 16 nm by 16 nm range.  All curves are offset for visual clarity.}
\label{fig:spectroscopy}
\end{figure}

We achieve dual control of the Fermi energy and perpendicular electric field by tuning the voltage on the silicon back gate and STM tip.  The total electric field arises from four different contributions: (1) the voltage applied to the back gate, (2) the voltage applied to the tip, (3) the fixed intrinsic charged impurities in the SiO$_2$, and (4) the work function difference between the graphene and the tungsten tip.  A finite electric field opens a band gap for the ABC-stacked trilayer graphene, as shown schematically in  Fig.~\ref{fig:schematic}(b), while the ABA-stacked trilayer remains metallic.   Figs.~\ref{fig:spectroscopy} (a) and (b) show dI/dV spectroscopy, which is proportional to the local density of states (LDOS), as a function of sample voltage and gate voltage for the ABA- and ABC-stacked trilayers respectively.  Each curve is the average of 64 measurements over a 16 nm by 16 nm region of the sample.  For both stacking orders, features move more positive in sample voltage as the gate voltage becomes more negative, due to the shifting of the Fermi energy.  However, the gate voltage also applies a larger perpendicular electric field as it becomes more negative.  For the ABA-stacked trilayer graphene in Fig.~\ref{fig:spectroscopy}(a), we see two peaks separated by roughly 30 meV (as well as a zero-bias anomaly dip which remains pinned to zero energy).  These peaks, marked by black and white arrows, move in parallel with changing gate voltage, which suggests they are relatively insensitive to the magnitude of the electric field at the densities we probe.  We calculate the band structure of ABA-stacked trilayer graphene using the low energy Hamiltonian~\cite{Peeters2009a,Peeters2009b} and find relatively good agreement for the evolution of our features with electric field.  Our calculation shows that while the single layer-like (roughly linear) bands move farther from the Fermi energy with increasing field, the bilayer-like (roughly quadratic) bands stay roughly evenly spaced and exhibit fairly flat band edges which result in our measured peaks (see Supplemental Material~\cite{SI}). 

For the remainder of this study, we focus on the spectroscopic results for the ABC-stacked trilayer graphene, where we open a field tunable band gap.  For the ABC-stacked trilayer graphene in Fig.~\ref{fig:spectroscopy}(b), we see a dip in the LDOS which grows in width as the gate tunes the center of the dip away from the Fermi energy (which occurs around $V_g$ = +25 V).  We attribute this dip to an electric field induced band gap (gap edges marked by black tick marks).  We also see a peak just above the conduction band edge which does not appear on the valence band side (marked with black arrows).  This peak is a signature of the van Hove singularity at the band edge, and grows in strength as the field becomes larger and the band becomes flatter.  This electron-hole asymmetry is present in all samples measured, but cannot be explained by the low energy Hamiltonian~\cite{Peeters2010}.

\begin{figure}[h]
\newpage
\includegraphics[width=8.6cm]{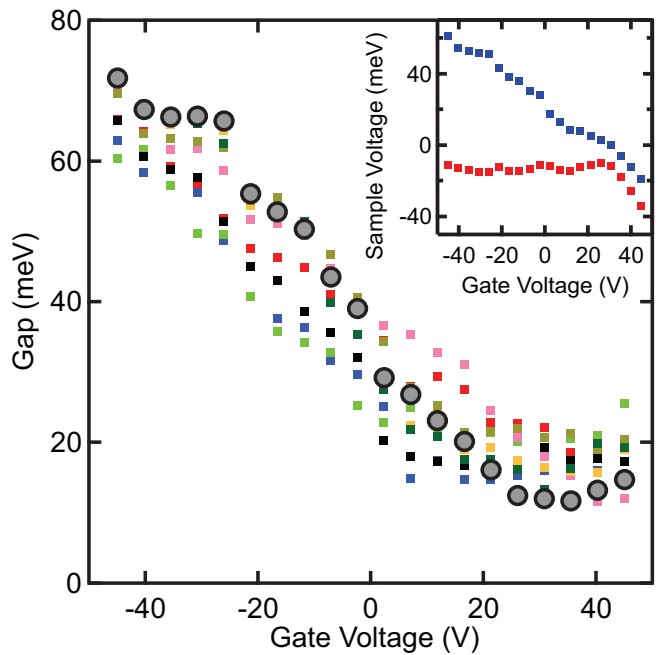} 
\caption{(Color online) Experimental band gap size in ABC trilayer graphene.  Gray circles represent the gap size for the data plotted in Fig. ~\ref{fig:spectroscopy}.  Colored squares represent the gap sizes for eight other spots on the sample.  Inset: Sample voltage of the valence (red) and conduction (blue) band edges as a function of back gate voltage for the gray circles.}
\label{fig:gaps}
\end{figure}

To extract band gap sizes, we adopt a fitting procedure in which we fit a V-shaped curve separated by a horizontal line to the region directly surrounding the minimum in the LDOS (see Supplemental Material~\cite{SI}).  The gap size is equal to the width of the horizontal line.  The gray circles in Fig. ~\ref{fig:gaps} show fits of the band gap from the curves of Fig. ~\ref{fig:spectroscopy}(b).  We measure an electric field induced gap as large as 70 meV at large negative back gate voltages and as small as 10 meV near the charge neutrality point (CNP).  The back gate serves to both tune the Fermi energy and apply an electric field, so the observed behavior is to be expected given that when the gate tunes the Fermi energy to the CNP, it also tunes the total electric field near its minimum.  Due to intrinsic sample doping, our minimum gap occurs at about +25 V on the back gate, but we can still see evidence of the gap growing on both sides of the CNP.  The inset of Fig.~\ref{fig:gaps} plots the energy of the conduction and valence band edges as a function of back gate voltage.  From this, as well as from Fig.~\ref{fig:spectroscopy}(b), we see that the majority of the gap opening is due to movement of the conduction band edge.  The valence band edge appears roughly pinned to zero sample voltage for back gate voltages far away from the CNP.  Both band edges move roughly in parallel when the back gate is tuned close to the CNP, as the gap is small there and the Fermi energy is no longer pinned near one band.  If we were able to tune our device to higher electron densities, we would expect to see the conduction band edge pinned near the Fermi energy. 

We repeat this analysis for eight other nearby spots on the sample.  Their respective gap sizes are shown by the colored squares in Fig. ~\ref{fig:gaps}.  At a given gate voltage, we see a substantial spatial variation in the size of the gap, on the order of about 20 meV.  When viewed against gate voltage, this variation is due to both the local shift in charge carrier density and electric field strength due to the underlying charge impurities.  

\begin{figure}[h]
\newpage
\includegraphics[width=8.6cm]{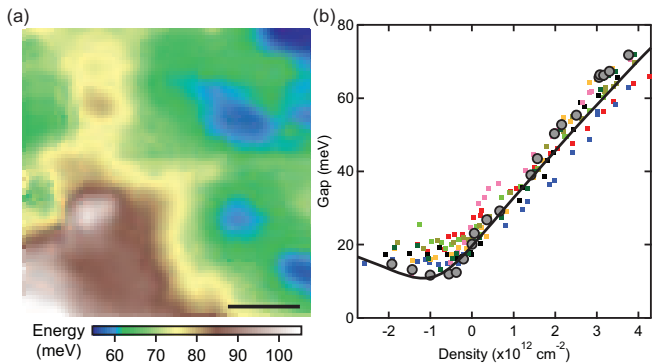} 
\caption{(color online) Charge fluctuation and experimental and theoretical band gap size in ABC trilayer graphene as a function of charge density. (a) Experimental dI/dV puddle map for ABC trilayer graphene taken at $V_g$ = +30 V.  The color scale represents the sample voltage of the van Hove singularity of the conduction band.  The scale bar is 10 nm. (b) Experimental gap size for ABC trilayer graphene as a function of charge density.  All color conventions are identical to Fig.~\ref{fig:gaps}.  The black curve is the theoretically expected gap size as a function of charge density.}
\label{fig:puddles}
\end{figure}

To quantify the effect of the trapped charge impurities on the local gap size, we record the LDOS with high spatial resolution.  For each spectroscopy curve, we determine the energy of a known feature of the band structure (we chose the van Hove singularity because it is very sharp and easy to fit, though we could equivalently chose the band gap center since these two features move roughly in parallel with changing charge density).  Fig.~\ref{fig:puddles}(a) shows the energy of the van Hove singularity for each spectroscopy curve in a 40 nm by 40 nm region of the sample with the Fermi energy tuned near the CNP.  Similar to the case of single~\cite{Yacoby2008,LeRoy2009a,Crommie2009} and bilayer graphene ~\cite{LeRoy2009b,Stroscio2011}, we notice puddles of charge on the sample which have no significant correlation with topographic features.  We measure a roughly Gaussian distribution with FWHM of 10.7 $\pm$ 0.8 meV.  Similar measurements on different spots on the same sample, and from different samples, give comparable charge variations.  We do not find significant changes in the charge variation as a function of back gate voltage, which is to be expected given that the low energy bands do not change in curvature much with changing energy and electric field.  By auto-correlating the charge puddle map, we are able to estimate a puddle size of 8 $\pm$ 0.2 nm for this sample spot.   

The presence of the puddles modifies the local charge density and hence the electric field at a given location.  For each of the locations in Fig.~\ref{fig:gaps}, we have found the local charge density due to the puddles.  Using this local charge density along with the charge density induced from the back gate, we are able to re-plot  the band gaps from Fig.~\ref{fig:gaps}  as a function of charge density instead of gate voltage.  The results are shown in Fig.~\ref{fig:puddles}(b).  This accounts for the charge fluctuations, and as a result the gap sizes from different spatial locations have much less variation.  The solid black line represents the theoretically expected band gap size as a function of charge density, calculated using the low energy Hamiltonian of ABC graphene which accounts for trigonal warping terms and interlayer screening~\cite{Peeters2010}.  We also add a constant charge density offset due to the work function mismatch between the tip and the trilayer graphene, which we estimate to be equivalent to -25 V on the back gate (see Supplemental Material~\cite{SI}).  When we account for this work function mismatch, we find theoretically that we are unable to close the band gap for any back gate voltage.  Furthermore, the minimum theoretical gap is larger than the expected zero-field gap of roughly 6 meV ~\cite{Lau2011} opened due to electron-electron interactions, therefore we safely ignore contributions to the band gap due to many-body effects.  We see very good agreement in both the slope and magnitude of our band gap at high carrier density, whereas at low density we tend to measure slightly larger than predicted gaps.  Low density gaps are likely larger than expected due to the residual charge fluctuations within each measurement region. 

We have presented local spectroscopic measurements of both the ABA- and ABC- stacked trilayer graphene systems.  We find that both interlayer screening and the relative work function difference between the top gate and graphene play an important role in determining the magnitude of the band gap in the ABC trilayer.  Finally, we see that local charge fluctuations strongly modify the size of the band gap and lead to spatially varying gap sizes.  For global electrical transport measurements, it will be critical to control these charge fluctuations in order to have a uniform band gap.    

We are thankful for discussions with Philippe Jacquod.  The work at Arizona was partially supported by the U. S. Army Research Laboratory and the U. S. Army Research Office under contract/Grant No. W911NF-09-1-0333 and the National Science Foundation EECS-0925152.  C.N.L. and F.W. are supported in part by NSF CAREER DMR/0748910, NSF/1106358, the FAME center, and UCOP.

\newpage

\setcounter{figure}{0}
\renewcommand{\thefigure}{S\arabic{figure}}

\section*{Supplementary Material}
\section{Calculation of ABA Band Structure}
We model ABA-stacked trilayer graphene as three coupled honeycomb lattices, each with two inequivalent lattice sites.  The layers have the standard Bernal-stacking scheme as described in Ref. ~\cite{Koshino2009}.  We account for intra-layer couplings between nearest neighbors, $\gamma_0$, coupling between inequivalent lattice sites lying directly above and below each other, $\gamma_1$, nearest-layer coupling between the next closest set of lattice sites, $\gamma_3$ and $\gamma_4$, next-nearest-layer coupling between equivalent lattice sites, $\gamma_2$ and $\gamma_5$, and the on-site energy difference between different layer lattice sites, $\Delta$ ~\cite{Koshino2009}.  We also account for the charge density on each layer given by $n_i$ induced via top ($n_t$) and back ($n_b$) gating leading to potential differences between the layers.  As such, we adopt the low energy Hamiltonian of Ref. ~\cite{Peeters2009}
\begin{widetext}
\begin{equation} \label{eq:ABA}
{H} = \left(
\begin{array}{cccccc}
-\Delta_{\rm 1,2}(n) + \Delta + \gamma_{\rm 5} & \gamma_ {\rm 0} f & \gamma_{\rm 1} & -\gamma_{\rm 4} f^* & \gamma_{\rm 5}/2 & 0 \\
\gamma_ {\rm 0} f^* & -\Delta_{\rm 1,2}(n) + \gamma_{\rm 2} & -\gamma_{\rm 4} f^* & \gamma_{\rm 3} f & 0 & \gamma_{\rm 2}/2 \\
\gamma_ {\rm 1} & -\gamma_{\rm 4} f & \Delta + \gamma_{\rm 5} & \gamma_{\rm 0} f^* & \gamma_{\rm 1} & -\gamma_{\rm 4} f \\
-\gamma_{\rm 4} f & \gamma_ {\rm 3} f^* & \gamma_{\rm 0} f & \gamma_{\rm 2} & -\gamma_{\rm 4} f & \gamma_{\rm 3} f^* \\
\gamma_{\rm 5}/2 &  0 & \gamma_{\rm 1} & -\gamma_{\rm 4} f^* & \Delta_{\rm 2,3}(n) + \Delta + \gamma_{\rm 5} & \gamma_{\rm 0} f \\
0 & \gamma_{\rm 2}/2 & -\gamma_{\rm 4} f^* & \gamma_{\rm 3} f & \gamma_{\rm 0} f^* & \Delta_{\rm 2,3}(n) + \gamma_{\rm 2} \\
\end{array}
\right) \, ,
\end{equation}
\end{widetext}
with $\Delta_{\rm 1,2}(n) = -\alpha |(n_{\rm 2} + n_{\rm 3} - n_{\rm b})|$ and  $\Delta_{\rm 2,3}(n) = -\alpha |(n_{\rm 3} - n_{\rm b})|$.  $f$ is defined as $f(k_{\rm x},k_{\rm_y}) = e^{i k_{\rm x} a_{\rm 0}/\sqrt3} + 2 e^{-i k_{\rm x} a_{\rm 0}/2\sqrt3} \cos{k_{\rm y} a_{\rm 0}/2}$, with $a_{\rm 0}$ = 2.46 \AA$ $  the length of the in-plane lattice vector.  We take $\alpha = e^2 c_{\rm 0} / \epsilon_{\rm 0} \kappa$ with $c_{\rm 0}$ = 3.35 \AA$ $ the interlayer distance and $\kappa$ = 2.3 the dielectric screening constant corresponding to graphene layers on SiO$_2$.  We take $\gamma_ {\rm 0}$ = 3.12 eV, $\gamma_ {\rm 1}$ = 0.377 eV, $\gamma_ {\rm 2}$ = -0.0206 eV, $\gamma_ {\rm 3}$ = 0.29 eV, $\gamma_ {\rm 4}$ = 0.12 eV, $\gamma_ {\rm 5}$ = 0.025 eV, and $\Delta$ = -0.009 eV ~\cite{Peeters2006}.  Our sign convention differs slightly from Ref. ~\cite{Peeters2009} for overall consistency.

We model our silicon back gate as a parallel plate capacitor capable of inducing a density $n_{\rm b} = \alpha_g (V_{\rm g} - V_{\rm 0})$, where $\alpha_g$ =  7.19 x 10$^{10}$cm$^{-2}$ V$^{-1}$ is determined by the gate capacitance with 300 nm of SiO$_2$, V$_ g$ is the applied gate voltage and V$_0$ is the shift of the charge neutrality point (CNP) due to intrinsic doping.  Similarly, we model our tip as a parallel plate capacitor with an effective distance of 1 nm.  We also add a constant density offset to the tip to account for the work function difference between the tip and trilayer graphene.  For ABA graphene and a tungsten tip, we estimate this to be equivalent to -10 V on the back gate, as this was the approximate difference between the global and local CNP.  Combining these effects, we estimate our tip induced density as a function of back gate induced density as $n_{\rm t} = -0.07 n_{\rm b} - \alpha_g V_{\rm offset}$, where $V_{\rm offset}$ is the effective back gate voltage due to the work function mismatch.  We use this, as well as the self-consistent screening calculations from Ref. ~\cite{Peeters2009}, to define the layer densities as $n_{\rm 2} = 0.3 (n_{\rm b} + n_{\rm t})$ and $n_{\rm 3} = 0.58 n_{\rm b} + 0.1 n_{\rm t}$.

\begin{figure}[b]
\includegraphics[width=8.6cm]{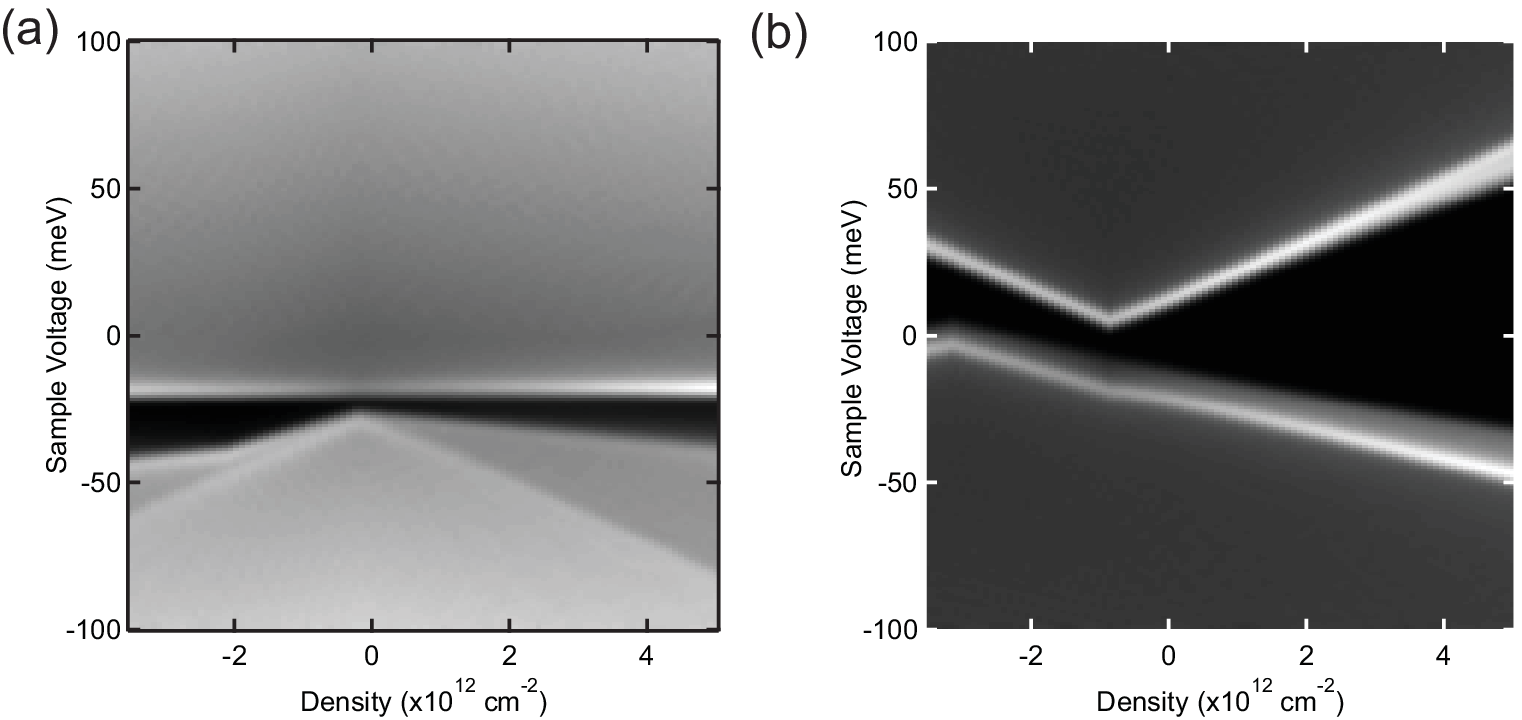} 
\caption{Density of states simulations for ABA and ABC trilayer graphene. (a) Calculated density of states for ABA trilayer graphene. (b) Calculated density of states for ABC trilayer graphene.  Darker color represents lower density of states.  Both calculations include a work function offset due to the tungsten STM tip.}
\label{fig:simulations}
\end{figure}

Fig. ~\ref{fig:simulations}(a) shows a numerical calculation of the density of states as a function of charge density, where darker colors represent lower density of states.  A linear shifting of the Fermi energy caused by the induced charge from the back gate is not included in this calculation.  The main features to notice are two spikes surrounding the minimum in the density of states.  These spikes originate from the van Hove singularities of the bilayer-like bands.  They are separated by about 25 meV, and their relative separation only weakly depends on the charge density.  We therefore attribute our two peaks in Fig. 2(b) of the main text to the van Hove singularities of the bilayer-like bands.  The dark region roughly surrounding -25 meV in sample voltage has a non-zero density of states, and thus we expect ABA trilayer graphene to remain metallic.

\section{Calculation of ABC Band Structure}\label{sec:pert}
Similarly to the case of ABA-stacked trilayer graphene, we model ABC-stacked trilayer graphene as three coupled honeycomb lattices, each with two inequivalent lattice sites.  The layers have the standard rhombohedral-stacking scheme as described in Ref. ~\cite{Peeters2010}. We account for the couplings described for the ABA-stacked trilayer appropriate for the ABC-stacked trilayer and adopt the low energy Hamiltonian of Ref. ~\cite{Peeters2010}
\begin{widetext}
\begin{equation} \label{eq:ABC}
{H} = \left(
\begin{array}{cccccc}
\Delta_{\rm 1,2}(n) & \gamma_ {\rm 0} f & \gamma_{\rm 1} & -\gamma_{\rm 4} f^* & 0 & 0 \\
\gamma_ {\rm 0} f^* & \Delta_{\rm 1,2}(n) & -\gamma_{\rm 4} f^* & \gamma_{\rm 3} f & 0 & \gamma_{\rm 2}/2 \\
\gamma_ {\rm 1} & -\gamma_{\rm 4} f & 0 & \gamma_{\rm 0} f & -\gamma_{\rm 4} f^* & \gamma_{\rm 3} f \\
-\gamma_{\rm 4} f & \gamma_ {\rm 3} f^* & \gamma_{\rm 0} f^* & 0 & \gamma_{\rm 1} & -\gamma_{\rm 4} f^* \\
0 &  0 & -\gamma_{\rm 4} f & \gamma_{\rm 1} & -\Delta_{\rm 2,3}(n) & \gamma_{\rm 0} f \\
0 & \gamma_{\rm 2}/2 & \gamma_{\rm 3} f^* & -\gamma_{\rm 4} f & \gamma_{\rm 0} f^* & -\Delta_{\rm 2,3}(n) \\
\end{array}
\right) \, ,
\end{equation}
\end{widetext}
where all parameters in Eqn. ~\ref{eq:ABC} are defined as in the prior section.

For our sample presented in the main text, we estimate the work function difference to be equivalent to -25 V on the back gate.  This is an educated guess, since we were unable to directly probe the global CNP.  However, prior ABC trilayer samples, where both the global and local CNPs were accessible, showed similar work function mismatches.  Furthermore, changing this offset corresponds to a small horizontal shift of the black curve from Fig. 4(b) of the main text, and does not qualitatively change our interpretation of the results.  Again, we use this along with the self-consistent screening calculations from Ref. ~\cite{Peeters2010} to define the layer densities as $n_{\rm 2} = 0.2 (n_{\rm b} + n_{\rm t})$ and $n_{\rm 3} = 0.6 n_{\rm b} + 0.2 n_{\rm t}$.

Fig. ~\ref{fig:simulations}(b) shows the same density of states calculation as in Fig. ~\ref{fig:simulations}(a), but for ABC-stacked trilayer graphene.  Again, the linear shifting of the Fermi energy due to the back gate is not included in the calculation.   The dark region roughly surrounding 0 meV in sample voltage has exactly zero density of states and represents the band gap.  We extract the theoretical magnitude of the indirect band gap as a function of charge density from this calculation.

\section{Fitting of Band Gaps}\label{sec:3}
We fit the band gap using a piecewise defined function given by
\begin{widetext}
\begin{equation} \label{eq:fiteqn}
f(x) = \begin{cases}
A(x-valence) + B & \text{for $x \le valence$} \\
	B & \text{for $valence < x < conduction$} \\
	C(x-conduction) + B & \text{for $x \ge conduction$}
\end{cases}
\end{equation}
\end{widetext}

\begin{figure}[b]
\includegraphics[width=8.6cm]{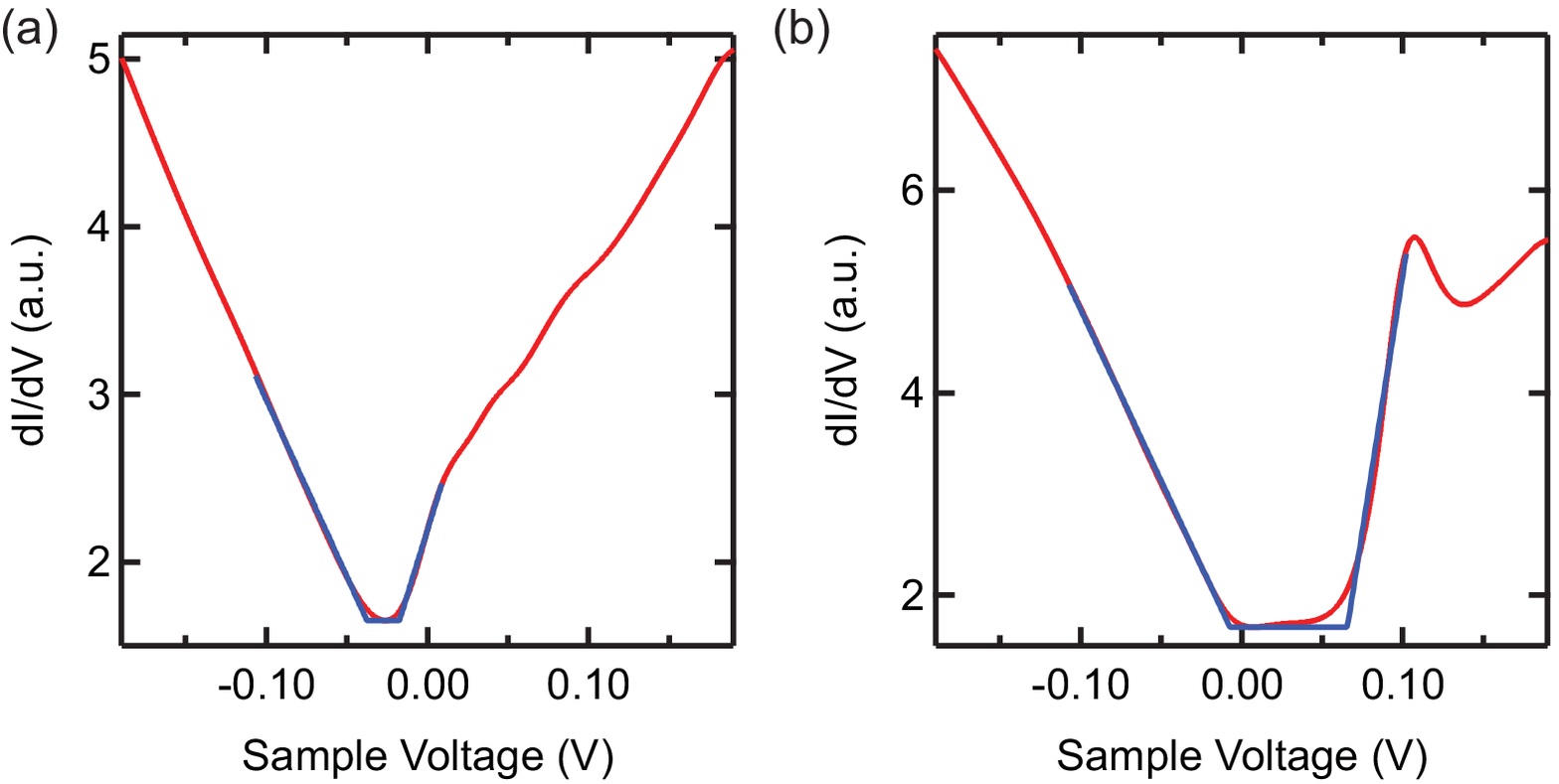} 
\caption{Example best fit gap determination curves at different back gate voltages. (a) The red curve is the average of 625 spectroscopy curves over a 50 nm by 50 nm region at V$_g$ = +45 V.  The blue curve is the best fit using our gap function and is only plotted over the fitted region of data. (b) Same as (a), but for V$_g$ = -45 V.}
\label{fig:gapfits}
\end{figure}

where $A$, $B$, $C$, $valence$, and $conduction$ are fit parameters.  The gap size is the energy separation between the $valence$ and $conduction$ parameters.  We also add the constraint that B must be equal to the minimum value of dI/dV (i.e. the gap minimum).  Finally, we only fit this function over a small range around the gap to avoid influences from the different slopes in dI/dV around the van Hove singularity in the conduction band side.  The red curve in Fig. ~\ref{fig:gapfits}(a) shows an example of a dI/dV curve at V$_g$ = +45 V averaged over a 50 nm by 50 nm region.  The blue curve is the best fit of the dI/dV curve using our gap fit function.  Fig ~\ref{fig:gapfits}(b) shows similar results for V$_g$ = -45 V.


\begin{thebibliography}{99}

\bibitem{Kim2012} K. S. Novoselov, V. I. Fal'ko, L. Colombo, P. R. Gellert, M. G. Schwab, and K. Kim,
{Nature} {\bf 490}, 192 (2012).

\bibitem{Novoselov2005} K. S. Novoselov, A. K. Geim, S. V. Morozov, D. Jiang, M. I. Katsnelson, I. V. Grigorieva, S. V. Dubonos, and A. A. Firsov,
{Nature} {\bf 438}, 197 (2005).

\bibitem{Zhang2005} Y. Zhang, Y.-W. Tan, H. L. Stormer, and P. Kim,
{Nature} {\bf 438}, 201 (2005).

\bibitem{CastroNeto2009} A. H. Castro Neto, F. Guinea, N. M. R. Peres, K. S. Novoselov, and A. K. Geim,  
{Rev. Mod. Phys.} {\bf 81}, 109 (2009).

\bibitem{DasSarma2011} S. Das Sarma, S. Adam, E. H. Hwang, and E. Rossi, 
{Rev. Mod. Phys.} {\bf 83}, 407 (2011).

\bibitem{Schweirz2012} F. Schwierz,
{Nat. Nanotechnol.} {\bf 5}, 487 (2010).

\bibitem{Guinea2006} F. Guinea, A. H. Castro Neto, and N. M. R. Peres, 
{Phys. Rev. B} {\bf 73}, 245426 (2006).

\bibitem{Aoki2007} M. Aoki and H. Amawashi, 
{Solid State Commun.} {\bf 142}, 123 (2007).

\bibitem{Koshino2009} M. Koshino, and E. McCann, 
{Phys. Rev. B} {\bf 79}, 125443 (2009).

\bibitem{Peeters2009a} A. A. Avetisyan, B. Partoens, and F. M. Peeters, 
{Phys. Rev. B} {\bf 79}, 035421 (2009).

\bibitem{Peeters2009b} A. A. Avetisyan, B. Partoens, and F. M. Peeters, 
{Phys. Rev. B} {\bf 80}, 195401 (2009).

\bibitem{Koshino2010} M. Koshino,
{Phys. Rev. B} {\bf 81}, 125304 (2010).

\bibitem{Kumar2011} S. B. Kumar and J. Guo,
{Appl. Phys. Lett.} {\bf 98}, 222101 (2011).

\bibitem{Wu2011} B.-R. Wu, 
{Appl. Phys. Lett.} {\bf 98}, 263107 (2011).

\bibitem{Tang2011} K. Tang, R. Qin, J. Zhou, H. Qu, J. Zheng, R. Fei, H. Li, Q. Zheng, Z. Gao, and J. Lu,
{J. Phys. Chem. C} {\bf 115}, 9458 (2011).

\bibitem{Craciun2009} M. F. Craciun, S. Russo, M. Yamamoto, J. B. Oostinga, A. F. Morpurgo, and S. Tarucha,
{Nat. Nanotechnol.} {\bf 4}, 383 (2009).

\bibitem{JH2011} T. Taychatanapat, K. Watanabe, T. Taniguchi, and P. Jarillo-Herrero, 
{Nat. Phys.} {\bf 7}, 621 (2011).

\bibitem{Henriksen2012} E. A. Henriksen, D. Nandi, , and J. P. Eisenstein, 
{Phys. Rev. X} {\bf 2}, 011004 (2012).

\bibitem{Peeters2010} A. A. Avetisyan, B. Partoens, and F. M. Peeters, 
{Phys. Rev. B} {\bf 81}, 115432 (2010).

\bibitem{MacDonald2010} F. Zhang, B. Sahu, H. Min, and A. H. MacDonald, 
{Phys. Rev. B} {\bf 82}, 035409 (2010).

\bibitem{Lau2011} W. Bao, L. Jing, J. Velasco Jr, Y. Lee, G. Liu, D. Tran, B. Standley, M. Aykol, S. B. Cronin, D. Smirnov, M. Koshino, E. McCann, M. Bockrath, and C. N. Lau,
{Nat. Phys.} {\bf 7}, 948 (2011).

\bibitem{Heinz2011} C. H. Lui, Z. Li, K. F. Mak, E. Cappelluti, and T. F. Heinz, 
{Nat. Phys.} {\bf 7}, 944 (2011).

\bibitem{Zhu2013} K. Zou, F. Zhang, C. Clapp, A. H. MacDonald, and J. Zhu,
{Nano Lett.} {\bf 13}, 369 (2013).

\bibitem{Lui2010} C. H. Lui, Z. Li, Z. Chen, P. V. Klimov, L. E. Brus, and T. F. Heinz,
{Nano Lett.} {\bf 11}, 164 (2011).

\bibitem{Zaliznyak2011} L. Zhang, Y. Zhang, J. Camacho, M. Khodas, and I. Zaliznyak, 
{Nat. Phys.} {\bf 7}, 953 (2011).

\bibitem{Cong2011} C. Cong, T. Yu, K. Sato, J. Shang, R. Saito, G. F. Dresselhaus, and M. S. Dresselhaus,
{ACS Nano} {\bf 5}, 8760 (2011).

\bibitem{SI} See Supplemental Material at [URL will be inserted by publisher] for calculations of trilayer band structures and explanation of band gap fitting procedure.

\bibitem{Yacoby2008} J. Martin, N. Akerman, G. Ulbricht, T. Lohmann, J. H. Smet, K. Von Klitzing, and A. Yacoby,
{Nature Phys.} {\bf 4}, 144 (2008).

\bibitem{LeRoy2009a} A. Desphande, W. Bao, F. Miao, C. N. Lau, and  B. J. LeRoy,
{Phys. Rev. B} {\bf 79}, 205411 (2009).

\bibitem{Crommie2009} Y. Zhang, V. W. Brar, C. Girit, A. Zettl, and M. F. Crommie, 
{Nat. Phys.} {\bf 5}, 722 (2009).

\bibitem{LeRoy2009b} A. Desphande, W. Bao, Z. Zhao, C. N. Lau, and B. J. LeRoy, 
{Appl. Phys. Lett.} {\bf 95}, 243502 (2009).

\bibitem{Stroscio2011} G. M. Rutter, S. Jung, N. N. Klimov, D. B. Newell, N. B. Zhitenev, and J. A. Stroscio,
{Nat. Phys.} {\bf 7}, 649 (2011).

\end{thebibliography}

\begin{thebibliography}{99}

\bibitem{Koshino2009} M. Koshino and E. McCann, 
{Phys. Rev. B} {\bf 79}, 125443 (2009).

\bibitem{Peeters2009} A. A. Avetisyan, B. Partoens, and  F. M. Peeters,
{Phys. Rev. B} {\bf 80}, 195401 (2009).

\bibitem{Peeters2006}  B. Partoens and F. M. Peeters, 
{Phys. Rev. B} {\bf 74}, 075404 (2006).

\bibitem{Peeters2010} A. A. Avetisyan, B. Partoens, and  F. M. Peeters,
{Phys. Rev. B} {\bf 81}, 115432 (2010).

\end{thebibliography}
\end{document}